\documentstyle[12pt]{article}

\newcounter{eq}
\newcounter{sc}

\def\overleftrightarrow#1{\vbox{\ialign{##\crcr
 $\leftrightarrow$\crcr\noalign{\kern-1pt\nointerlineskip}
 $\hfil\displaystyle{#1}\hfil$\crcr}}}

\newlength{\minitwocolumn}
\begin{document}

\begin{flushright}
DPUR/TH/51\\
August, 2016\\
\end{flushright}
\vspace{20pt}

\pagestyle{empty}
\baselineskip15pt

\begin{center}
{\large\bf Reissner-Nordstrom Solution from Weyl Transverse Gravity
\vskip 1mm }

\vspace{20mm}
Ichiro Oda \footnote{E-mail address:\ ioda@phys.u-ryukyu.ac.jp
}

\vspace{5mm}
           Department of Physics, Faculty of Science, University of the 
           Ryukyus,\\
           Nishihara, Okinawa 903-0213, Japan.\\

\end{center}


\vspace{5mm}
\begin{abstract}
We study classical solutions in the Weyl-transverse (WTDiff) gravity coupled to an electro-magnetic
field in four space-time dimensions. The WTDiff gravity is invariant under both the local Weyl (conformal) 
transformation and the volume preserving diffeormorphisms (transverse diffeomorphisms) and is known to be 
equivalent to general relativity at least at the classical level (perhaps even in the quantum regime). 
In particular, we find that only in four space-time dimensions, the charged Reissner-Nordstrom black hole 
metric is a classical solution when it is expressed in the Cartesian coordinate system.
\end{abstract}

\newpage
\pagestyle{plain}
\pagenumbering{arabic}


\rm
\section{Introduction}

It is fair to say that the success of modern particle physics has been 
achieved with the help of the concept of spontaneous symmetry breakdown of
local gauge symmetries \cite{Nambu}. However, gravity is outside the purview 
of this success, and what was worse, one veritable crisis is generated
once gravity is introduced into the standard model of particle physics.
This crisis is nowdays called the cosmological constant problem \cite{Weinberg}, 
and in the context at hand it might be simply phrased as follows: 
Astronomical observation shows that the cosmological constant at present is 
many orders of magnitude smaller than that estimated in modern particle physics.

The situation associated with the cosmological constant problem becomes worse
when quantum effects are taken into account. Namely, every time the higher-order
loop corrections are added in perturbation theory, we need to fine-tuning
the value of the cosmological constant. The cosmological constant problem is
therefore one of the most difficult hierarchy problems.
 
A number of theoretical physicists have advocated changing the rules of general relativity
in such a way that the cosmological constant emerges as an integration constant
which is irrelevant to any parameters in the action. This approach is sometimes called 
unimodular gravity \cite{Einstein}-\cite{Padilla}, and it has been expected that unimodular
gravity could suppress the radiative corrections coming from the higher-order loop corrections in 
perturbation theory although the value of the cosmological constant must be fixed by an intial 
condition. \footnote{Recently, we have established a topological model where the Newton's constant is determined 
by an initial condition \cite{Oda3}-\cite{Oda5}.}
   
Recently, a new type of unimodular gravity, which is called the Weyl-transverse (WTDiff) gravity
\cite{Izawa}-\cite{Oda1}, has been developed 
where as a new ingredient, the local Weyl (conformal) symmetry is introduced into the original unimodular 
gravity, which has only the tranverse diffeomorphisms (TDiff), or equivalently, the volume-preserving diffeomorphisms
as local symmetries. The Weyl symmetry is known to be the original symmetry of the conventional gauge symmetries
and is expected to help us to understand the short distance phenomena for a long time \cite{tHooft}.
Furthermore, in order to realize the Weyl symmetry in a gravitational theory without introducing higher-derivative 
terms, we usually need a scalar field, and just the existence of the scalar field provides us with 
a possibility such that gravity is also unified in the standard model via spontaneous symmetry breakdown
of the Weyl symmetry. 

One of the purposes in this article is to study classical solutions in the WTDiff gravity. Even if the WTDiff
gravity is equivalent to general relativity at the classical level (perhaps, even at the quantum level
\cite{Englert}-\cite{Benedetti}), to our knowledge, nobody has explicitly derived classical solutions 
within the framework of the WTDiff gravity except our recent work of the Schwarzschild black hole metric \cite{Oda2}.

In particular, we wish to investigate whether the charged Reissner-Nordstrom black hole metric is included 
in classical solutions when the WTDiff gravity is coupled to an electro-magnetic field. In reality, a highly 
charged black hole would be quickly neutralized by interactions with matter in its vicinity and therefore 
such a solution is not extremely relevant to realistic astrophysical situations. Nevertheless, charged black
holes illustrate a number of important features of more general situations \cite{Caroll}, so it is worthwhile
to verify its existence in classical solutions of the WTDiff gravity coupled to an electro-magnetic field. 

This paper is organised as follows: In Section 2, we review the WTDiff gravity. To get the WTDiff gravity, 
we start with the conformally invariant scalar-tensor gravity, and then fix the Weyl symmetry by a gauge condition 
which does not violate the TDiff \cite{Oda2}. In Section 3, we solve the equations of motion of the WTDiff gravity
coupled to an electro-magnetic field in the static and spherically symmetric ansatz. We show that the Reissner-Nordstrom 
metric in the Cartesian coordinate system is in fact a classical solution. The final section is devoted to discussions.

\section{Brief review of the Weyl-transverse (WTDiff) gravity}

We will start with an action of the conformally invariant scalar-tensor gravity in a general $n$ dimensional space-time 
\footnote{We follow notation and conventions by Misner et al.'s textbook \cite{MTW}, for instance, 
the flat Minkowski metric $\eta_{\mu\nu} = diag(-, +, +, +)$, the Riemann curvature tensor 
$R^\mu \, _{\nu\alpha\beta} = \partial_\alpha \Gamma^\mu_{\nu\beta} - \partial_\beta \Gamma^\mu_{\nu\alpha} 
+ \Gamma^\mu_{\sigma\alpha} \Gamma^\sigma_{\nu\beta} - \Gamma^\mu_{\sigma\beta} \Gamma^\sigma_{\nu\alpha}$, 
and the Ricci tensor $R_{\mu\nu} = R^\alpha \, _{\mu\alpha\nu}$.
The reduced Planck mass is defined as $M_p = \sqrt{\frac{c \hbar}{8 \pi G}} = 2.4 \times 10^{18} GeV$.
Throughout this article, we adopt the reduced Planck units where we set $c = \hbar = M_p = 1$.
In this units, all quantities become dimensionless. 
Finally, note that in the reduced Planck units, the Einstein-Hilbert Lagrangian density takes the form
${\cal L}_{EH} = \frac{1}{2} \sqrt{-g} R$.} \footnote{This conformally invariant gravity theory has a wide application 
in phenomenology and cosmology \cite{Oda6}-\cite{Oda9}.}

\begin{eqnarray}
S = \int d^n x \ \sqrt{-g} \left[ \frac{n-2}{8(n-1)} \varphi^2 R +  \frac{1}{2}
g^{\mu\nu} \partial_\mu \varphi \partial_\nu \varphi  \right],
\label{Cof-inv S-T Action 1}
\end{eqnarray}
which is invariant under not only general coordinate transformation but also Weyl transformation of 
the metric tensor $g_{\mu\nu}$ and a ghost-like scalar field $\varphi$ as 
\begin{eqnarray}
g_{\mu\nu} \rightarrow g^\prime_{\mu\nu} = \Omega^2(x) g_{\mu\nu}, \quad
\varphi \rightarrow \varphi^\prime = \Omega^{- \frac{n-2}{2}}(x) \varphi,
\label{Weyl transf}
\end{eqnarray}
where $\Omega(x)$ is an arbitrary scalar function of space-time coordinates $x^\mu$.

With a gauge condition for the Weyl transformation
\begin{eqnarray}
\varphi = 2 \sqrt{\frac{n-1}{n-2}},
\label{Gauge 1}
\end{eqnarray}
the conformally invariant scalar-tensor gravity produces the Einstein-Hilbert action of general relativity. 
On the other hand, a gauge condition for the longitudinal diffeomorphism
\begin{eqnarray}
\varphi = 2 \sqrt{\frac{n-1}{n-2}} |g|^{- \frac{n-2}{4n}},
\label{Gauge 2}
\end{eqnarray}
leads to an action of the WTDiff gravity
\begin{eqnarray}
S = \frac{1}{2} \int d^n x \ |g|^{\frac{1}{n}} \left[ R + \frac{(n-1)(n-2)}{4n^2} \frac{1}{|g|^2}
g^{\mu\nu} \partial_\mu |g| \partial_\nu |g|  \right],
\label{WTDiff Action 1}
\end{eqnarray}
where we have defined $g = \det g_{\mu\nu} < 0$. The action (\ref{WTDiff Action 1}) turns out to be 
invariant under not the full group of diffeomorphisms (Diff) but only the transverse diffeomorphisms (TDiff) 
in addition to the Weyl transformation \cite{Oda2}. 
From this derivation, the WTDiff gravity is found to be at least classically equivalent to general relativity 
since the both actions are obtained via different choices of gauge condition from the same action 
(\ref{Cof-inv S-T Action 1}).  We conjecture that this equivalence holds even in the quantum regime 
\cite{Englert}-\cite{Benedetti}. To put differently, we believe that there are no anomalies associated with 
the Weyl symmetry and the longitudinal diffeomorphism.

Next, we will derive the equations of motion for the WTDiff gravity (\ref{WTDiff Action 1}).
A method of the derivation is to work with the action (\ref{Cof-inv S-T Action 1}) of the
conformally invariant scalar-tensor gravity, derive its equations of motion, and then
substitute the gauge condition (\ref{Gauge 2})
into them. After some calculations, it turns out that the action (\ref{Cof-inv S-T Action 1}) produces 
the equations of motion for $g_{\mu\nu}$ and $\varphi$, respectively 
\begin{eqnarray}
\frac{n-2}{8(n-1)} \left[ \varphi^2 G_{\mu\nu} + ( g_{\mu\nu} \Box - \nabla_\mu \nabla_\nu ) 
(\varphi^2)\right] = \frac{1}{4} g_{\mu\nu} \partial_\rho \varphi \partial^\rho \varphi 
- \frac{1}{2} \partial_\mu \varphi \partial_\nu \varphi,
\label{Eq of motion of conf-ST 1}
\end{eqnarray}
and 
\begin{eqnarray}
\frac{n-2}{4(n-1)} \varphi R = \Box \varphi,
\label{Eq of motion of conf-ST 2}
\end{eqnarray}
where $G_{\mu\nu} = R_{\mu\nu} - \frac{1}{2} g_{\mu\nu} R$ is the Einstein tensor 
and $\Box \varphi = g^{\mu\nu} \nabla_\mu \nabla_\nu \varphi$. It is easy to show that
the equation of motion (\ref{Eq of motion of conf-ST 2}) for the $\it{spurion}$ field $\varphi$ is
not an independent equation, but can be derived from the trace part of Eq. (\ref{Eq of motion of conf-ST 1}).

Then, substituting the gauge condition (\ref{Gauge 2}) into Eq. (\ref{Eq of motion of conf-ST 1}) gives us 
the equations of motion for the WTDiff gravity 
\begin{eqnarray}
R_{\mu\nu} - \frac{1}{n} g_{\mu\nu} R = T_{\mu\nu} - \frac{1}{n} g_{\mu\nu} T,
\label{Eq of motion of WTDiff}
\end{eqnarray}
where the energy-momentum tensor $T_{\mu\nu}$ is defined as
\begin{eqnarray}
T_{\mu\nu} = \frac{(n-2)(2n-1)}{4n^2} \frac{1}{|g|^2} \partial_\mu |g| \partial_\nu |g|
- \frac{n-2}{2n} \frac{1}{|g|} D_\mu D_\nu |g|,
\label{Energy-momentum}
\end{eqnarray}
where we have defined $D_\mu D_\nu |g| = \partial_\mu \partial_\nu |g| - \Gamma^\rho_{\mu\nu} \partial_\rho |g|$.
Note that Eq. (\ref{Eq of motion of WTDiff}) is purely the traceless part of the Einstein field equations. 
By an explicit calculation, it is easy to verify that the equations of motion (\ref{Eq of motion of WTDiff})
are invariant under both the Weyl transformation and the TDiff \cite{Oda2}. 

Since the whole formalism, that is, the conformally invariant scalar-tensor gravity, is generally covariant,
the energy-momentum tensor (\ref{Energy-momentum}) should satisfy the covariant conservation law
\begin{eqnarray}
\nabla^\mu T_{\mu\nu} = 0.
\label{Consv-Energy-momentum}
\end{eqnarray}
Of course we can verify the validity of Eq. (\ref{Consv-Energy-momentum}) by a direct calculation,
and its calculation is simplified by going to a local Lorentz frame where the affine connection and
first derivatives of the metric tensor vanish. Taking the covariant derivative of 
Eq. (\ref{Eq of motion of WTDiff}) and using the Bianchi identity 
\begin{eqnarray}
\nabla^\mu ( R_{\mu\nu} - \frac{1}{2} g_{\mu\nu} R ) = 0,
\label{Bianchi}
\end{eqnarray}
produces 
\begin{eqnarray}
\frac{n-2}{2n} \nabla_\mu R = - \frac{1}{n} \nabla_\mu T.
\label{Cov-derivative}
\end{eqnarray}
This equation means that $R + \frac{2}{n-2} T$ is a constant, which we will call $\frac{2n}{n-2} \Lambda$
\begin{eqnarray}
R + \frac{2}{n-2} T = \frac{2n}{n-2} \Lambda.
\label{Lambda}
\end{eqnarray}
Eq. (\ref{Eq of motion of WTDiff}), together with Eq. (\ref{Lambda}), yields the Einstein equations
\begin{eqnarray}
R_{\mu\nu} - \frac{1}{2} g_{\mu\nu} R + \Lambda g_{\mu\nu} = T_{\mu\nu}.
\label{Einstein equations}
\end{eqnarray}
Although we have recovered the Einstein equations from the equations of motion of the WTDiff gravity,
the cosmological constant $\Lambda$ appears as a mere integration constant and has nothing to do with
any terms in the action or vacuum fluctuations. In other words, Eq. (\ref{Eq of motion of WTDiff}) does
not essentially include the cosmological constant and the contribution from the radiative
corrections to the cosmological constant cancels in the RHS of Eq. (\ref{Eq of motion of WTDiff}),
thereby guaranteeing the stability of the cosmological constant against the quantum corrections.
This phenomenon is the most appealing point of the WTDiff gravity and unimodular gravity.

\section{Reissner-Nordstrom solution}

In a recent article \cite{Oda2}, we have investigated classical solutions in the WTDiff gravity and found that
the Schwarzschild metric is indeed a classical solution to the equations of motion of the WTDiff gravity, 
(\ref{Eq of motion of WTDiff}). A study of the Schwarzschild solution is of importance since the Schwarzschild solution 
corresponds to the basic one-body problem of classical astronomy, and the reliable experimental verifications 
of the Einstein equations are almost based on the Schwarzschild line element.
Then, it is natural to ask ourselves whether the charged Reissner-Nordstrom black hole metric is also a classical solution 
to the equations of motion of the WTDiff gravity coupled to an electro-magnetic field or not.
In this section, we will see that it is indeed the case.

Before attempting to show that the charged Reissner-Nordstrom metric is a classical solution,
we must recall that an action of a $U(1)$ gauge field is invariant under the Weyl transformation only in four space-time 
dimensions, so henceforth we will confine ourselves to the specific case of $n = 4$.  Then, an action of the WTDiff gravity
coupled to an electro-magnetic field $A_\mu$ reads
\begin{eqnarray}
S = \int d^4 x \ \left[ \frac{1}{2} |g|^{\frac{1}{4}} \left( R + \frac{3}{32} \frac{1}{|g|^2}
g^{\mu\nu} \partial_\mu |g| \partial_\nu |g|  \right) 
- \frac{1}{4} |g|^{\frac{1}{2}} g^{\mu\nu} g^{\rho\sigma} F_{\mu\rho} F_{\nu\sigma} \right],
\label{WTDiff Action plus EM}
\end{eqnarray}
where we have defined the field strength of the $U(1)$ gauge field by $F_{\mu\nu} = \partial_\mu A_\nu - \partial_\nu A_\mu$.

From the action (\ref{WTDiff Action plus EM}), it is straightforward to derive the equations of motion for
the metric tensor $g_{\mu\nu}$ and the gauge field $A_\mu$. The result for the metric variation is given by
\begin{eqnarray}
R_{\mu\nu} - \frac{1}{4} g_{\mu\nu} R = T_{\mu\nu} - \frac{1}{4} g_{\mu\nu} T,
\label{Metric eq of WTDiff plus EM}
\end{eqnarray}
where the energy-momentum tensor $T_{\mu\nu}$ is defined as
\begin{eqnarray}
T_{\mu\nu} = \frac{7}{32} \frac{1}{|g|^2} \partial_\mu |g| \partial_\nu |g|
- \frac{1}{4} \frac{1}{|g|} D_\mu D_\nu |g| + |g|^{\frac{1}{4}} F_{\mu\alpha}
F_\nu \, ^\alpha.
\label{4d energy-momentum}
\end{eqnarray}
And the Maxwell equations take the form 
\begin{eqnarray}
\partial_\nu ( \sqrt{-g} F^{\mu\nu} ) = 0.
\label{Maxwell eq}
\end{eqnarray}

Now we wish to look for a gravitational field outside an isolated, static, spherically symmetric object
with mass $M$ and electric charge $Q$. In the far region from the isolated object, we assume that the metric tensor 
is in an asymptotically Lorentzian form
\begin{eqnarray}
g_{\mu\nu} \rightarrow \eta_{\mu\nu} + {\cal {O}}\left(\frac{1}{r}\right),
\label{BC}
\end{eqnarray}
where $\eta_{\mu\nu}$ is the Minkowski metric, and the radial coordinate $r$ is defined as 
\begin{eqnarray}
r = \sqrt{(x^1)^2 + (x^2)^2 + (x^3
)^2} = \sqrt{(x^i)^2},
\label{radial}
\end{eqnarray}
with $i$ running over spatial coordinates ($i = 1, 2, 3$).

Let us recall that the most spherically symmetric line element in four space-time dimensions 
reads \cite{Oda2} 
\begin{eqnarray}
d s^2 = - A(r) d t^2 + (d x^i)^2 + B(r) (x^i d x^i)^2,
\label{Line element 1}
\end{eqnarray}
where $A(r)$ and $B(r)$ are functions depending on only $r$.

From this line element (\ref{Line element 1}), we can read off the non-vanishing components of 
the metric tensor 
\begin{eqnarray}
g_{tt} = - A, \quad  g_{ij} = \delta_{ij} + B x^i x^j,
\label{Metric}
\end{eqnarray}
and the components of its inverse matrix are 
\begin{eqnarray}
g^{tt} = - \frac{1}{A}, \quad  g^{ij} = \delta^{ij} - \frac{B}{1 + B r^2} x^i x^j.
\label{Inverse Metric}
\end{eqnarray}
Moreover, using these components of the metric tensor, the affine connection is calculated to be
\begin{eqnarray}
\Gamma^t_{ti} &=& \frac{A^\prime}{2A} \frac{x^i}{r}, \quad 
\Gamma^i_{tt} = \frac{A^\prime}{2(1 + B r^2)} \frac{x^i}{r}, \nonumber\\ 
\Gamma^i_{jk} &=& \frac{1}{2(1 + B r^2)} \frac{x^i}{r} ( 2B r \delta_{jk} + B^\prime x^j x^k ),
\label{Affine}
\end{eqnarray}
where the primes on $A(r)$ and $B(r)$ denote the differentiation with respect to $r$, for instance,
$A^\prime = \frac{dA}{dr}$. 

As for the electro-magnetic field $A_\mu(x)$, we assume that it has a static, spherically symmetric
electric potential,
\begin{eqnarray}
A_t = - \phi (r), \quad  A_i = 0,
\label{Gauge field}
\end{eqnarray}
where $\phi (r)$ is a function of $r$. 

Next, in order to show that the Reissner-Nordstrom line element is a classical solution to the field equations
(\ref{Metric eq of WTDiff plus EM}) and (\ref{Maxwell eq}), we would like to take a gauge condition 
for the Weyl transformation
\begin{eqnarray}
g = -1,
\label{g=-1}
\end{eqnarray}
which is nothing but the unimodular condition so that this gauge condition does not break the transverse diffeomorphisms
(TDiff). The reason why we pick up this gauge condition (\ref{g=-1}) is the following: Look at our Einstein equations 
(\ref{Metric eq of WTDiff plus EM}) where the energy-momentum tensor $T_{\mu\nu}$ receives the contribution of
the determinant of the metric tensor in addition to the electro-magnetic $U(1)$ gauge field. The Reissner-Nordstrom metric 
is a solution of the Einstein equations in the vacuum in the sense that the energy-momentum tensor $T_{\mu\nu}$ is vanishing 
except the part of the electro-magnetic field. The gauge condition (\ref{g=-1}) then satisfies this requirement. 

With the metric tensor ansatz (\ref{Line element 1}), the gauge condition (\ref{g=-1}) is reduced to the form
\begin{eqnarray}
A (1 + B r^2) = 1.
\label{AB}
\end{eqnarray}
Using this gauge condition (\ref{AB}) and Eqs. (\ref{Metric})-(\ref{Affine}), the Ricci tensor
and the scalar curvature can be easily calculated to be  
\begin{eqnarray}
R_{tt} &=& \frac{1}{2} A ( A^{\prime\prime} + \frac{2}{r} A^\prime ), \nonumber\\ 
R_{ij} &=& \left[ \frac{1}{r^2} ( 1 - A ) - \frac{A^\prime}{r} \right] \delta_{ij}
+ \frac{1}{r^2} \Biggl[ \frac{1}{r^2} ( A - 1 ) 
+ \frac{1}{r} \frac{A^\prime}{A} ( A - 1 ) - \frac{1}{2} \frac{A^{\prime\prime}}{A} \Biggr] 
x^i x^j,
\nonumber\\
R &=& - A^{\prime\prime} - \frac{4}{r} A^\prime - \frac{2}{r^2} ( A - 1 ). 
\label{Curvature}
\end{eqnarray}
These results are used to evaluate  the non-vanishing components of the {\it{traceless}} Einstein 
tensor defined as $G_{\mu\nu}^T \equiv R_{\mu\nu} - \frac{1}{4} g_{\mu\nu} R$
\begin{eqnarray}
G_{tt}^T &=& \frac{1}{4} A \left[ A^{\prime\prime} - \frac{2}{r^2} ( A - 1 ) \right], \nonumber\\ 
G_{ij}^T &=& \frac{1}{4} \left( \delta_{ij} - \frac{A + 1}{A} \frac{x^i x^j}{r^2} \right) 
\left[ A^{\prime\prime} - \frac{2}{r^2} ( A - 1 ) \right].
\label{G^T}
\end{eqnarray}

First, let us solve the Maxwell equations (\ref{Maxwell eq}). With the ansatzes (\ref{Line element 1})
and (\ref{Gauge field}) and the gauge condition (\ref{AB}), the Maxwell equations (\ref{Maxwell eq})
are cast to a single equation
\begin{eqnarray}
\frac{d}{dr} ( r^2 \phi^\prime ) = 0,
\label{Maxwell eq 2}
\end{eqnarray}
which is easily integrated to be
\begin{eqnarray}
\phi(r) = \frac{\sqrt{2} Q}{r} + c,
\label{phi}
\end{eqnarray}
where $Q$, which corresponds to an electric charge, and $c$ are integration constants. To fix the constant
$c$, we will impose a boundary condition
\begin{eqnarray}
\lim_{r \rightarrow \infty} \phi(r) = 0,
\label{BC for phi}
\end{eqnarray}
which determines $c = 0$. Thus, we obtain the final expression for $\phi(r)$
\begin{eqnarray}
\phi(r) = \frac{\sqrt{2} Q}{r}.
\label{phi 2}
\end{eqnarray}

Next, let us try to solve the {\it{traceless}} Einstein equations (\ref{Metric eq of WTDiff plus EM}). 
For this purpose, we will calculate the {\it{traceless}} energy-momentum tensor defined as 
$T_{\mu\nu}^T \equiv T_{\mu\nu} - \frac{1}{4} g_{\mu\nu} T$ whose result is summarized as
\begin{eqnarray}
T_{tt}^T &=& A \frac{Q^2}{r^4}, \nonumber\\ 
T_{ij}^T &=& \left( \delta_{ij} - \frac{A + 1}{A} \frac{x^i x^j}{r^2}  \right) \frac{Q^2}{r^4}. 
\label{T^T}
\end{eqnarray}

Consequently, the {\it{traceless}} Einstein equations (\ref{Metric eq of WTDiff plus EM}) reduce to 
an equation
\begin{eqnarray}
A^{\prime\prime} - \frac{2}{r^2} ( A - 1 ) = \frac{4 Q^2}{r^4},
\label{A-eq}
\end{eqnarray}
which can be rewritten in a more manageable form 
\begin{eqnarray}
\frac{1}{r} \frac{d^2}{d r^2} \left[ r \left( A - 1 - \frac{Q^2}{r^2} \right) \right]
- \frac{2}{r^2} \frac{d}{d r} \left[ r \left( A - 1 - \frac{Q^2}{r^2} \right) \right] = 0.
\label{A-eq 2}
\end{eqnarray}
By performing an integration, $A(r)$ is given by
\begin{eqnarray}
A(r) = 1 - \frac{2M}{r} + \frac{Q^2}{r^2} + a r^2,
\label{A-value}
\end{eqnarray}
where $M$ and $a$ are integration constants. From the boundary condition (\ref{BC}),
we have to choose $a = 0$, and we can obtain the expression for $B(r)$ in terms of the gauge
condition (\ref{AB}). As a result, we reach the expressions for $A(r)$ and $B(r)$ 
\begin{eqnarray}
A(r) = 1 - \frac{2M}{r} + \frac{Q^2}{r^2}, \quad B(r) = \frac{2Mr - Q^2}{r^2 (r^2 - 2Mr + Q^2)}.
\label{AB-value}
\end{eqnarray}
Then, the line element is of form
\begin{eqnarray}
d s^2 = - \left( 1 - \frac{2M}{r} + \frac{Q^2}{r^2} \right) d t^2 + (d x^i)^2 
+ \frac{2Mr - Q^2}{r^2 (r^2 - 2Mr + Q^2)} (x^i d x^i)^2.
\label{RN 1}
\end{eqnarray}
Accordingly, we have succeeded in showing that the Reissner-Nordstrom metric in the Cartesian coordinate system
is a classical solution in the WTDiff gravity coupled to the Maxwell field.

Here we should refer to an important remark. The Reissner-Nordstrom metric (\ref{RN 1}) in the Cartesian coordinate 
system can be rewritten in a more familiar form in the spherical coordinate system
\begin{eqnarray}
d s^2 = - \left( 1 - \frac{2M}{r} + \frac{Q^2}{r^2} \right) d t^2 
+ \frac{1}{1 - \frac{2M}{r} + \frac{Q^2}{r^2}} d r^2 + r^2 d \Omega^2,
\label{RN 2}
\end{eqnarray}
where 
\begin{eqnarray}
d \Omega^2 = d \theta^2 + \sin^2 \theta d \varphi^2.
\label{Omega}
\end{eqnarray}
However, this expression (\ref{RN 2}) is {\it{not}} a classical solution in  the WTDiff gravity plus 
the Maxwell field. This is a notable feature of the WTDiff gravity where there is no the full group
of diffeomorphisms, but only the TDiff.

\section{Discussions}

In this article, it is shown that the Weyl-transverse (WTDiff) gravity plus the Maxwell action has the charged Reissner-Nordstrom
black hole metric as a classical solution when the metric is expressed in the Cartesian coordinate system.
It is of interest to note that the conventional form of the Reissner-Nordstrom metric in the spherical coordinate
system is not a classical solution. The same statement is also valid in case of the Schwarzshild black hole metric
presented in our previous article \cite{Oda2}. This dependence on the coordinate systems of the classical solutions 
in the WTDiff gravity is a characteristic feature since the TDiff are defined as a subgroup of the full diffeomorphisms
such that the determinant of the transformation matrix is the unity
\begin{eqnarray}
J \equiv \det J^\alpha_{\mu \prime} \equiv \det \frac{\partial x^\alpha}{\partial x^{\mu \prime}} = 1.
\label{TDiff}
\end{eqnarray}
When we transform the Reissner-Nordstrom metric from the Cartesian coordinate system to the spherical one,
we encounter the non-trivial vaule of $J$ 
\begin{eqnarray}
J = r^2 \sin \theta,
\label{J}
\end{eqnarray}
by which the Reissner-Nordstrom metric in the spherical coordinate system is not a solution. Actually, if
we rewrite a flat Minkowski space-time in the spherical coordinate system, the resultant line element is not
a classical solution owing to the non-trivial Jacobian factor even if it is a classical solution in the
Cartesian coordinate (Here we assume $Q = 0$). 

It is straightforward to generalize the Reissner-Nordstrom black hole solution to a more general static, 
spherically symmetric solution where a black hole carries both electric and magnetic charge.
A nontrivial generalization is to obtain the Reissner-Nordstrom black hole solution in arbitrary space-time
dimensions except four dimensions since the conventional Maxwell action is not invariant under the Weyl transformation
in such dimensions. Another interesting problem is to look for classical solutions with the property of
$g \neq -1$. We wish to return to these problems in future.

\begin{flushleft}
{\bf Acknowledgements}
\end{flushleft}
This work is supported in part by the Grant-in-Aid for Scientific 
Research (C) No. 16K05327 from the Japan Ministry of Education, Culture, 
Sports, Science and Technology.



\begin{thebibliography}{99}

\bibitem{Nambu}
Y. Nambu and G. Jona-Lasinio, {Phys. Rev. {\bf 122} (1961) 345.}

\bibitem{Weinberg}
S. Weinberg, {Rev. Mod. Phys. {\bf 61} (1989) 1.}

\bibitem{Einstein}
A. Einstein, {in {"The Principle of Relativity", by A. Einstein et al., Dover
Publications, New York, 1952.}}

\bibitem{Anderson}
J. L. Anderson and D. Finkelstein, {Am. J. Phys. {\bf 39} (1971) 901.}

\bibitem{Bij}
J. van der Bij, H. van Dam and Y. J. Ng, {Physica {\bf 116A} (1982) 307.}

\bibitem{Buchmuller1}
W. Buchmuller and N. Dragon, {Phys. Lett. {\bf B 207} (1988) 292.}

\bibitem{Henneaux}
M. Henneaux and C. Teitelboim, {Phys. Lett. {\bf B 222} (1989) 195.}

\bibitem{Buchmuller2}
W. Buchmuller and N. Dragon, {Phys. Lett. {\bf B 223} (1989) 313.}

\bibitem{Unruh}
W. G. Unruh, {Phys. Rev. {\bf D 40} (1989) 1048.}

\bibitem{Ng}
Y. J. Ng and H. van Dam, {J. Math. Phys. {\bf 32} (1991) 1337.}

\bibitem{Faedo1}
E. Alvarez and A. F. Faedo, {Phys. Rev. {\bf D 76} (2007) 064013.}

\bibitem{Faedo2}
E. Alvarez, A. F. Faedo and J. J. Lopez-Villarejo, {JHEP {\bf 0810} (2008) 023.}

\bibitem{Ellis}
G. F. R. Ellis, H. van Elst, J. Murugan and J. -P. Uzan, {Class. Quant. Grav. 
{\bf 28} (2011) 225007.}

\bibitem{Padilla}
A. Padilla and I. D. Saltas, {Eur. Phys. J. {\bf C 75} (2015) 561.}

\bibitem{Oda3}
I. Oda, {Adv. Studies in Theor. Phys. {\bf 10} (2016) 319, arXiv:1602.00851 [gr-qc].}

\bibitem{Oda4}
I. Oda, {arXiv:1602.03478 [gr-qc], PTEP (in press).}

\bibitem{Oda5}
I. Oda, {arXiv:1603.00112 [gr-qc], Int. J. Mod. Phys. D (in press).}

\bibitem{Izawa}
K-I. Izawa, {Prog. Theor. Phys {\bf 93} (1995) 615.}

\bibitem{Barcelo1}
C. Barcelo, R. Carballo-Rubio and L. J. Garay, {Phys. Rev. {\bf D 89} (2014) 124019.}

\bibitem{Alvarez1}
E. Alvarez, S. Gonzalez-Martin, M. Herrero-Valea and C. P. Martin, {JHEP {\bf  1508} (2015) 078.}

\bibitem{Alvarez2}
E. Alvarez, S. Gonzalez-Martin, M. Herrero-Valea and C. P. Martin, {Phys. Rev. {\bf D92} (2015) 061502.}

\bibitem{Alvarez3}
E. Alvarez, S. Gonzalez-Martin, M. Herrero-Valea and C. P. Martin, {Phys. Rev. {\bf D93} (2016) 123018.}

\bibitem{Alvarez4}
E. Alvarez, S. Gonzalez-Martin and C. P. Martin, {arXiv:1604.07263 [hep-th].}

\bibitem{Oda1}
I. Oda, {arXiv:1606.01571 [gr-qc].}

\bibitem{tHooft}
G. 't Hooft, {arXiv:1410.6675 [gr-qc].}

\bibitem{Englert}
F. Englert, C. Truffin and R. Gastmans, {Nucl. Phys. {\bf B 117} (1976) 407.}

\bibitem{Shaposhnikov1}
M. Shaposhnikov and D. Zenhausern, {Phys. Lett. {\bf B 671} (2009) 162.}

\bibitem{Percacci}
R. Percacci, {New Jour. of Phys. {\bf 13} (2011) 125013.}

\bibitem{Monin}
F. Gretsch and A. Monin, {Phys. Rev. {\bf D 92} (2015) 045036.}

\bibitem{Carballo-Rubio}
R. Carballo-Rubio, {Phys. Rev. {\bf D 91} (2015) 124071.}

\bibitem{Benedetti}
D. Benedetti, {Gen. Rel. Grav. {\bf 48} (2016) 68.}

\bibitem{Oda2}
I. Oda, {arXiv:1607.06562 [gr-qc].}

\bibitem{Caroll}
S. M. Caroll, {"Spacetime and Geometry", Addison Wesley, 2004.}

\bibitem{MTW}
C. W. Misner, K. S. Thorne and J. A. Wheeler, {"Gravitation", W H Freeman and Co (Sd), 1973.}

\bibitem{Oda6}
I. Oda, {Phys. Rev. {\bf D 87} (2013) 065025, arXiv:1301.2709 [hep-ph].}

\bibitem{Oda7}
I. Oda, {Phys. Lett. {\bf B 724} (2013) 160, arXiv:1305.0884 [hep-ph].}

\bibitem{Oda8}
I. Oda, {Adv. Studies in Theor. Phys. {\bf 8} (2014) 215, arXiv:1308.4428 [hep-ph].}

\bibitem{Oda9}
I. Oda and T. Tomoyose, {JHEP {\bf 09} (2014) 165, arXiv:1407.7575 [hep-th].}





 
\end{thebibliography}
\end{document}